\documentclass[10pt,twoside]{epnt1p}
\usepackage{graphicx}

\usepackage{amssymb}
\usepackage{amsmath}

\usepackage{url}

\newcommand{\lsim}{\raisebox{-0.13cm}{~\shortstack{$<$ \\[-0.07cm] $\sim$}}~}
\newcommand{\gsim}{\raisebox{-0.13cm}{~\shortstack{$>$ \\[-0.07cm] $\sim$}}~}

\begin{document}

\begin{frontmatter}


\begin{flushright}
{\small UG-FT-212/06, CAFPE-82/06}
\end{flushright}

\title{Probing TeV gravity at neutrino telescopes}


\author[address1]{J.I. Illana}, 
\author[address1]{M. Masip},    
\author[address2]{D. Meloni}    

\address[address1]{CAFPE and Depto.~de F{\'\i}sica Te\'orica y del Cosmos, U. de Granada, 18071 Granada, Spain}
\address[address2]{INFN and Dipto.~di Fisica, Universit\`a degli Studi di Roma ``La Sapienza", 00185 Rome, Italy}

\begin{abstract}
Models with extra dimensions and the fundamental scale at the TeV could imply signals in large neutrino telescopes due to gravitational scattering of cosmogenic neutrinos in the detection volume. Apart from the production of microscopic black holes, extensively studied in the literature, we present 
gravity-mediated interactions at larger distances, that can be calculated in the eikonal approximation. In these elastic processes the neutrino loses a small fraction of energy to a hadronic shower and keeps going. The event rate of these events is higher than that of black hole formation and the signal is distinct: no charged leptons and possibly multiple-bang events.
\end{abstract}


\end{frontmatter}


\section{Motivation: cosmogenic neutrinos and TeV gravity}

Cosmogenic neutrinos, produced in the scattering of protons off cosmic microwave background photons, have access to TeV physics in interactions with terrestrial nucleons at center of mass energies $\sqrt{s}=\sqrt{2m_NE_\nu}\gsim10$~TeV. If the fundamental scale of gravity is $M_D\sim1$~TeV \cite{ADD}, which may happen in $D>4$ spacetime dimensions, these $\nu N$ interactions are transplanckian, $\sqrt{s}>M_D$.

The only consistent theory known so far in such a regime, string theory, tells us that the interactions are soft in the ultraviolet. The scattering amplitudes vanish except in the forward region, an effect that can be understood as the destructive interference of string excitations \cite{SR}. The forward amplitudes are dominated by the zero mode of the string, corresponding to the exchange of a gauge particle of spin 1, ${A}\sim gs/t$, for open strings, or a graviton of spin 2, ${A}\sim (1/M^2_D) s^2/t$, for closed strings. Therefore, one expects that gravity dominates in transplanckian collisions.

It must be noticed that present bounds on $M_D$ from colliders (LEP, Tevatron) \cite{colliders} or astrophysics and cosmology (supernovae cooling) \cite{SN} come from processes at energies below $M_D$ and are indirect, since they actually constrain the energy emitted to Kaluza-Klein gravitons of mass $M\propto R^{-1}$ in the $n=D-4$ compact extra dimensions, which is a function of the compactification radius $R$. Those bounds rely on the assumption of all extra dimensions being large (the effective and fundamental Planck scales then relate through $M_P^2\propto R^n M_D^{2+n}$) that can be evaded in more sophisticated compactification models \cite{Giudice:2004mg}. In contrast, transplanckian collisions probe $M_D$ directly and independently of compactification details.

\section{Gravitational interactions}

We have shown that gravitational interactions are the only relevant in the transplanckian regime of energies. In impact parameter space, one must keep in mind two critical values: the Planck length $\lambda_D\sim M_D^{-1}$ and the Schwarzschild radius $R_S(s)\sim(\sqrt{s}/M_D)^{1/(n+1)}M_D^{-1}$. There are two types of interactions.

Short-distance interactions, with impact parameter $b\lsim R_S$, in which the colliding particles (a neutrino and a parton inside the nucleon) collapse into a black hole (BH) correspond to the exchange of strongly coupled gravitons of high momentum (non-linear gravity). The collapse involves strongly coupled gravity and is not calculable perturbatively. Most analyses are based on a geometric cross section \cite{BH} $\hat\sigma_{\rm BH}\simeq\pi R_S^2(\hat s)$ for the partonic process, with $\hat s=xs$. If $\sqrt{\hat s}\gg M_D$, namely $R_S\gg M_D^{-1}$, one expects that this estimate will not be off by any large factors \cite{Giddings:2004xy}. However, most of the BHs produced in the scattering of an ultrahigh energy neutrino off a parton are light, with masses just above $M_D$, since the $\nu N$ cross section is dominated by the low $x$ region. In this regime the amount of gravitational radiation emitted during the collapse or the topology of the singularity are important effects that add uncertainty to the geometric estimate.

Long-distance interactions, with $b\gg R_S$, have to do with the exchange of weakly coupled gravitons of low momentum (linearized gravity) \cite{eik}. In transplackian collisions quantum gravity acts inside the event horizon ($R_S>\lambda_D$). Therefore, these elastic interactions are due to classical gravity. They are characterized by a small deflection angle $\theta^*$ in the center of mass (CM) frame, 
\begin{eqnarray}
\theta^*\sim\frac{\sqrt{\hat s}}{M_D^{n+2}b^{n+1}}
\sim\left(\frac{R_S}{b}\right)^{n+1}\ll 1\Rightarrow
y=q^2/\hat s=\frac{1}{2}(1-\cos\theta^\star)\ll 1\ .
\end{eqnarray}
The elastic collision of a neutrino and a parton that exchange $D$-dimensional gravitons is then described by the {\em eikonal} amplitude resumming an infinite set of ladder and cross-ladder diagrams in the limit in which the momentum $q$ carried by each graviton is smaller than the CM energy or, in terms of the fraction of energy lost by the incoming neutrino, $y=(E_\nu-E'_\nu)/E_\nu\ll1$.
In this limit the amplitude is independent of the spin of the colliding particles.
Essentially, ${A}_{\rm eik}$ is the exponentiation of the Born
amplitude in impact parameter space \cite{riccardo}:
\begin{eqnarray}
{A}_{\rm eik}(\hat s,t)=\frac{2 \hat s}{i}
\int {\rm d}^2b\; e^{i\mathbf{q}\cdot\mathbf{b}}\;
\left(e^{i\chi (\hat s,b)}-1\right)
\equiv 4\pi\hat sb_c^2F_n(b_cq)\ ,
\end{eqnarray}
where $\chi (\hat s,b)$ is the eikonal phase,
\begin{eqnarray}
\chi(\hat s,b) = \frac{1}{2s}\int\frac{{\rm d}q}{(2\pi)^2}
{\rm e}^{-i{\bf q}\cdot{\bf b}}{A}_{\rm Born}(\hat s,q^2)
\equiv \left(\frac{b_c}{b}\right)^n
\end{eqnarray}
and a new scale $b_c$ appears,
\begin{eqnarray}
b_c(\hat s)=\left[\displaystyle\frac{(4\pi)^{\frac{n}{2}-1}}{2}\Gamma\left(
\displaystyle\frac{n}{2}\right)\frac{\hat s}{M_D^{n+2}}\right]^\frac{1}{n}\ .
\end{eqnarray}
The total partonic cross sections can be obtained from the amplitudes above using the optical theorem. They are $\hat\sigma_{\rm eik}\propto b_c^2\sim \hat s^\frac{2}{n}$, growing faster with energy than the BH cross sections $\hat\sigma_{\rm BH}\propto R_S^2\sim \hat s^\frac{1}{n+1}$.

Therefore, in transplanckian collisions one may consider two types of processes \cite{us}: elastic (long-distance) {\em soft} processes where the neutrino transfers to the partons a small fraction $y<y_{\rm max}$ of its energy and keeps going, and shorter distance ($b<R_S$) {\em hard} processes where the neutrino loses in the collision most of its energy, possibly collapsing into a BH. We take $y_{\rm max}=0.2$, the typical inelasticity of a standard model (SM) interaction, but any value of the order of 0.1 yields similar results. On the other hand, there is a $y_{\rm min}=E_{\rm thres}/E_\nu$ determined by the threshold energy $E_{\rm thres}$ transfered to a parton that produces an observable hadronic cascade.
The corresponding $\nu N$ cross sections are obtained by convolution with the parton distribution functions $f_{\{q,\bar q,g\}}(x,\mu)$ with the appropriate energy scale $\mu$ \cite{riccardo,us}. 

To estimate the relative frequency of both type of processes \cite{us}, consider a  $10^{10}$~GeV neutrino that scatters off a nucleon with $E_{\rm thres}=100$~TeV and $M_D=1$~TeV for $n=2\;(6)$ extra dimensions. The number of eikonal interactions before the neutrino gets destroyed is the ratio of interaction lengths $L_{\rm BH}/L_{\rm eik}=\sigma_{\rm eik}/\sigma_{\rm BH}=12.5\;(1.64)$. For a SM interaction $L_{\rm SM}=440$~km while $L_{\rm BH}=17$~km (4 km) in ice. The total energy lost by the neutrino in these eikonal interactions and the energy lost to graviton radiation are relatively small: $E^{\rm loss}_{\rm eik}=5.9\times10^7$~GeV ($1.2\times10^8$~GeV) and  $E_{\rm loss}^{\rm rad}=9.2\times10^7$~GeV ($1.2\times10^8$~GeV) in 1 km of ice.

\section{Signals at neutrino telescopes}

The flux of cosmogenic neutrinos, yet unobserved, depends on the production rate of primary nucleons of energy around and above the GZK cutoff. It is correlated with proton and photon fluxes that must be consistent, respectively, with the number of ultrahigh energy events at AGASA and HiRes \cite{Anchordoqui:2002hs} and with the diffuse $\gamma$-ray background measured by EGRET \cite{Sreekumar:1997un}. We base our analysis on the two neutrino fluxes described in \cite{semikoz}. The first one saturates the observations by EGRET, whereas for the second one the correlated flux of $\gamma$-rays contribute only a 20\% to the data, with the nucleon flux normalized in both cases to AGASA/HiRes. The {\it higher} flux predicts 820 downward neutrinos of each flavor with energy between $10^8$~GeV and $10^{11}$~GeV per year and km$^2$, versus 370 for the {\it lower} one. The spectrum has a peak at neutrino energies between $10^9$~GeV and $10^{10}$~GeV.

When a neutrino hits a nucleon it will start a hadronic shower. The total number of hadronic events in a neutrino telescope of cross sectional area $A$ in a time $T$ is
\begin{eqnarray}
N_{\rm events}=2\pi AT\int {\rm d}E_\nu\sum_{\nu_i,\bar \nu_i} \frac{{\rm d}\phi_{\nu_i}}{{\rm d}E_\nu}\int {\rm d}\cos\theta_z P_{\rm surv}P_{\rm int}\ ,
\end{eqnarray}
where $P_{\rm surv}$ is the probability that the neutrino survives to reach the detector and $P_{\rm int}$ the probability that it interacts inside the detector:
\begin{eqnarray}
P_{\rm surv}(E_\nu,\theta_z)&=&{\rm e}^{-x(\theta_z)N_A(\sigma_{\rm SM}+\sigma_{\rm BH})}\ , \quad
P_{\rm int}(E_\nu)\approx 1-{\rm e}^{-L\rho_{\rm ice} N_A\sigma^{\nu N}_{\rm int}}\ ,
\end{eqnarray}
with $x$ the column density of material, $\theta_z$ the zenith angle and $L$ the longitudinal detector size. When $L$ is larger than the interaction length $L_0$, there may be multiple-bang events. Neglecting the energy lost by the neutrino in each interaction, the probability of $N$ bangs and the average, and most probable, number of bangs are, respectively
\begin{eqnarray}
P_N(L)&=&{\rm e}^{-L/L_0}\frac{(L/L_0)^N}{N!}\ ,\quad 
\langle N\rangle=\sum_{N=1}^\infty NP_N=L/L_0\ .
\end{eqnarray}

\begin{figure*}[t]
\includegraphics[width=0.32\linewidth]{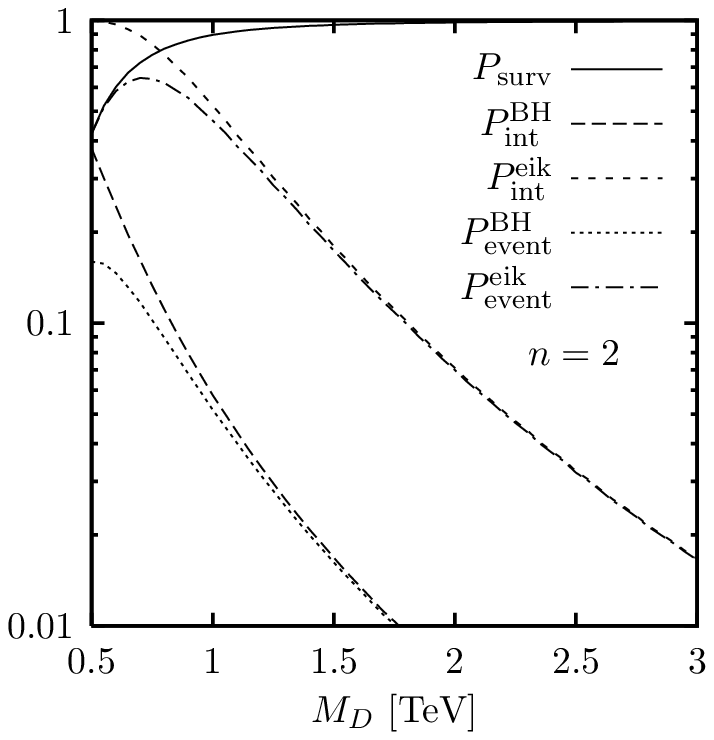}
\hfill
\includegraphics[width=0.32\linewidth]{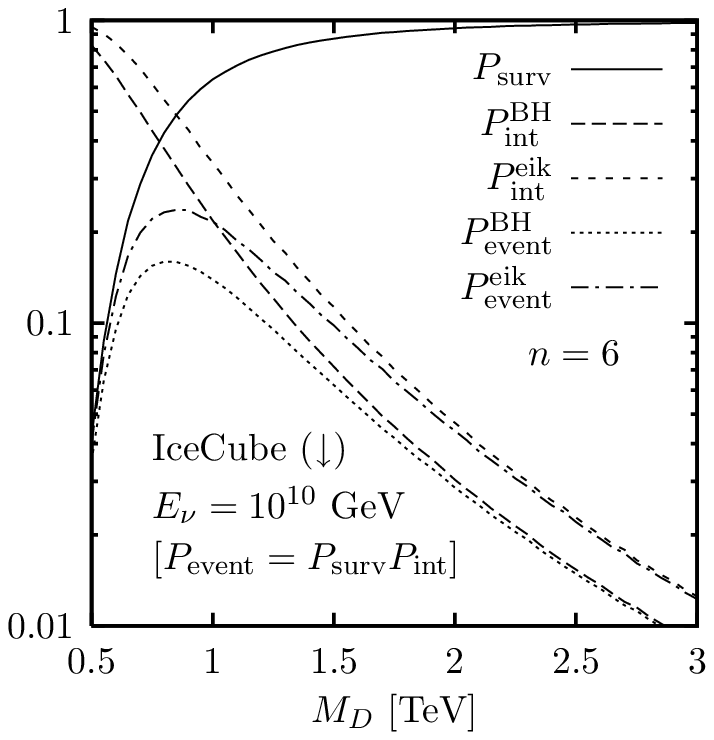}
\hfill
\includegraphics[width=0.32\linewidth]{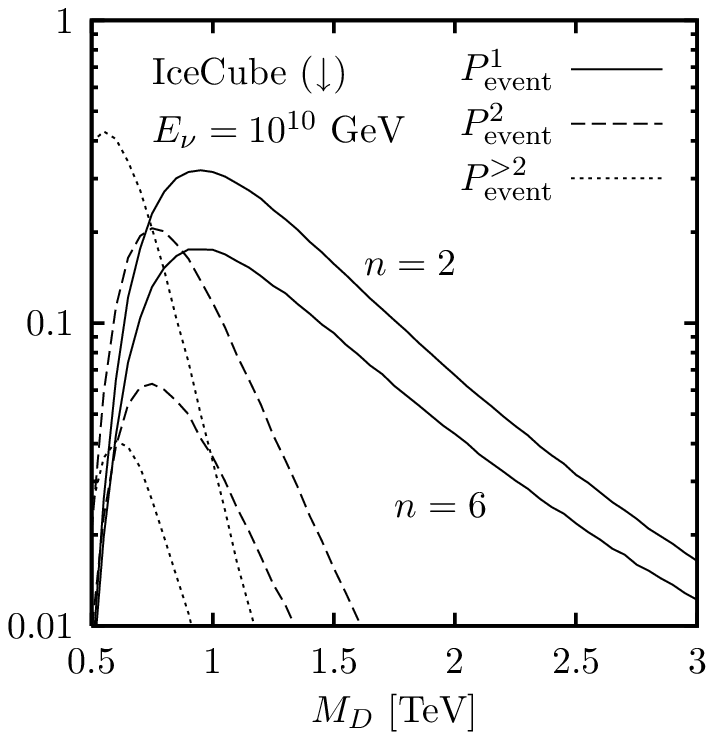}
\caption{
Probabilities defined in the text for a $10^{10}$~GeV neutrino reaching IceCube from $\theta_z=0$ as a function of $M_D$ for $n=2$ and $n=6$.}
\label{fig1}
\end{figure*}
The different probabilities for a typical cosmogenic neutrino of $10^{10}$~GeV reaching vertically IceCube are shown in Fig.~\ref{fig1} for illustration. Double-bang events could also be produced by SM interactions (the decay of a tau created in a first interaction) or in the BH evaporation. For the double-bang tau event to be contained inside a detector like IceCube (1 km of length with 125 m between strings), the energy of the tau lepton must be between $2.5\times 10^{6}$ GeV and $10^{7}$ GeV. In this case, the probability is only $6.8\times10^{-5}$.

\begin{figure*}[t]
\begin{minipage}[t]{0.48\linewidth}
\includegraphics[width=\linewidth]{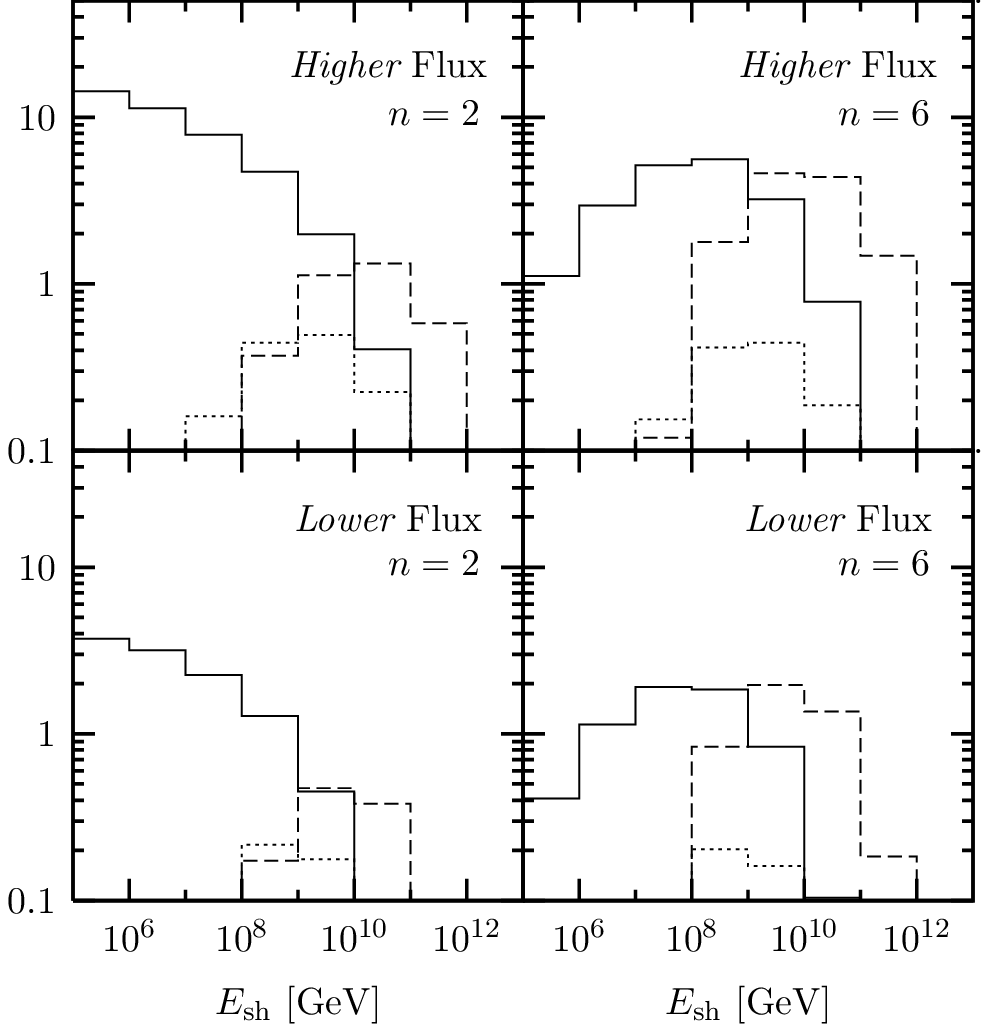}
\caption{
Energy distribution (events per bin) of the eikonal (solid), BH (dashed)
and SM (dotted) events in IceCube per year for the {\it higher} and
the {\it lower}
cosmogenic fluxes, $M_D=2$~TeV and $n=2,\ 6$.}
\label{fig2}
\end{minipage}\hfill
\begin{minipage}[t]{0.48\linewidth}
\includegraphics[width=\linewidth]{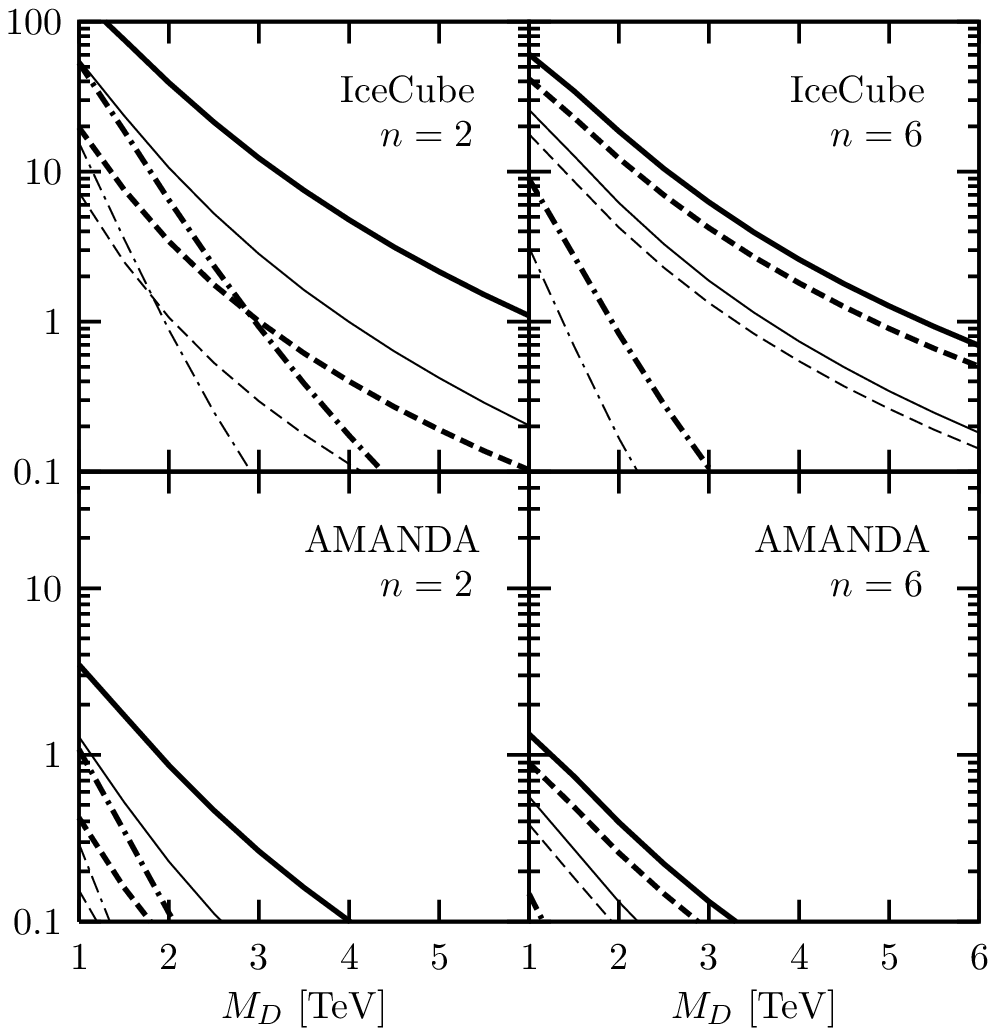}
\caption{
Contained events per year in IceCube and AMANDA for the
{\it higher} (thick) and the {\it lower} (thin) cosmogenic
fluxes and $n=2,\ 6$.
We show eikonal (solid), multi-bang (dashed-dotted) and BH (dashed)
events.}
\label{fig3}
\end{minipage}
\end{figure*}

The energy distribution of the hadronic cascades and the total number of black hole and eikonal events at AMANDA (0.03 km$^2$ and a length of 700 m) and IceCube (1 km$^3$) for the neutrino fluxes introduced above are given in Figs.~\ref{fig2} and \ref{fig3}, respectively. In the SM we expect 1.32 (0.50) contained events per year in IceCube for the higher (lower) flux. Of those, 0.38 (0.14) would come from a neutral current and 0.94 (0.36) from a charged current.

\section{Conclusions}

Cosmogenic neutrinos directly probe TeV gravity in transplackian collisions with nucleons at Earth if there are $D>4$ spacetime dimensions. Two types of interactions take place. Hard processes, when the impact parameter is smaller than the Schwarzschild radius of the neutrino-parton system, produce mostly light black holes with theoretically uncertain cross sections. At larger distances, soft elastic processes occur in which the neutrinos lose a small fraction of energy to a hadronic shower, in a well known regime described by the eikonal approximation.

The latter turn out to be dominant and produce a clear signal in large neutrino telescopes: contained hadronic showers without charged leptons. Furthermore, this signal cannot be confused with ordinary SM events due to an unexpectedly high neutrino flux because in the SM 24\% of the events are accompanied by muons and multiple-bang events are very suppressed, in contrast to the elastic gravitational events. The values of the fundamental scale of gravity that IceCube could reach are comparable to those to be explored at the LHC.


\setcounter{section}{0}
\setcounter{subsection}{0}
\setcounter{figure}{0}
\setcounter{table}{0}
\newpage
\end{document}